\documentclass[floatfix,showpacs,amsmath,amssymb,letterpaper,groupaddresses,superscriptaddress]{article}
\usepackage{latexsym}
\usepackage{epsfig}

\usepackage{amsmath}
\usepackage{amssymb}
\usepackage{graphicx}
\usepackage{times}
\usepackage[section]{placeins}
\usepackage{caption}
\usepackage{subcaption}

\usepackage{color}

\def\be{\begin{equation}}
\def\ee{\end{equation}}
\def\ba{\begin{eqnarray}}
\def\ea{\end{eqnarray}}

\def\ra{\rangle}

\def\h{\hskip 1cm}

\usepackage[symbol]{footmisc}

\begin{document}

\vspace{4cm}
\begin{center}{\Large \bf Encoding the information in relative parameters}\\
\vspace{2cm}
\vspace{2cm}

F . Rezazadeh$^1$\h A . Mani$^{2}$\footnote[2]{mani.azam@ut.ac.ir}\\
\vspace{1cm} $^1$ Department of Physics, Sharif University of Technology, P.O. Box 11155-9161, Tehran, Iran.\\
$^2$ Department of Engineering Science, Collage of Engineering, University of Tehran, Tehran, Iran.\\

\vskip 2cm

\begin{abstract}
We investigate the problem of communicating three parameters in the absence of shared reference frame.
We explore two methods in which the relative angles of spins are used to encode the parameters. In the first method we use three spins that carry the information in their relative angles while in the second method we use three disjoint spin-pairs and the information is sent through the relative angles of each individual pair. We show that in the first method, the information conveyed by each qubit is more than the second one, and that is while it requires fewer particles.

\end{abstract}
\end{center}


\vskip 2cm

\section{Introduction}
Quantum information transmission protocols are regarded as specific classes of multipartite tasks and in almost all these protocols it is implicitly assumed that the parties agree on an external reference frame, with respect to which the information is encoded and measurements are performed  \cite{quantum teleportation, dense coding, key distribution1, key distribution2, QKD3, QKD4}. In the absence of such external Shared Reference Frame (SRF), a natural question is that with respect to the local coordinate of which party, the protocols should be run, i.e. the operators, observations and measurements should be formulated. To face the problem, one strategy is using some (classical or quantum) resources to set a common reference frame, while the other strategy is using reference independent methods of quantum communication.
The first approach (Establishing a Shared Reference Frame) has been studied from different points of view \cite{ Gisin and Popescu, Began, Peres and Scudo, S.n, optimal measurment1, optimal measurment2, optimal measurment3, Massar} and it is known that establishing a perfect SRF requires transmission of infinite amount of information \cite{Massar and Popescu, our paper}.
On the other hand any finite (i.e. imperfect) SRF must be treated quantum mechanically and naturally disturbs during measurements and this degradation should be considered in the implementation of quantum protocols \cite{degradation}.\\
 
Alternatively, the second strategy includes different reference independent methods of information sharing \cite{DFS1, DFS2,classical and quantum communication,communication without SRF 2, Rudolph, keydistributionNOSRF1, keydistributionNOSRF2, keydistributionNOSRF3, keydistributionNOSRF4}.
The most well-known method in this context is the approach of Decoherence Free Subspaces (DFS)  \cite{DFS1, DFS2}. Decoherence free subspace is a set of quantum states which are insensitive to some special noise. In the absence of a SRF, decoherence free subspace is the set of states which are invariant under any arbitrary global rotation. Using the techniques of DFS method, the authors of \cite{classical and quantum communication} have shown that classical and quantum communication is possible without a shared reference frame. 
There are still many other works that investigate the problem of communication without SRF in the context of DFS methods \cite{communication without SRF 1, communication without SRF 3, SRFapproach1}.\\

In this paper we use another strategy for communication without SRF. Briefly stated, in DFS-based methods, the information is encoded in  \textit{particular states} of the system which are invariant under global rotations but we encode the information in \textit{particular parameters} of the system which are invariant under global rotations. It means that, although the state of the system changes under global rotation but those specific parameters are rotationally invariant and consequently have the same value in all reference frames. 
In fact, some parameters of a quantum system can be  regarded to be relative or collective \cite{Rudolph}. The collective parameters describe the system with respect to something external and hence they change with a global rotation. The relative parameters describe parts of the system with respect to each other and so they are invariant under a global rotation. For example, in a two-qubit system, the individual coordinates of each qubit with respect to an external  coordinate are collective parameters, while the angle between the spins is a relative one. \\


To shed light on the importance of the relative parameters in the absence of a SRF, consider the case that Alice wishes to send some information to Bob by sending a two-qubit system. If Alice and Bob have a SRF, Alice can encode the information in both relative or collective parameters of the system. 
On the other hand, in the absence of a SRF, Alice can not use the individual orientations of the spins with respect to her own frame to perfectly communicate with Bob, but she can surely use the relative angle of the spins to send the information since this angle is independent of any external frame \cite{Rudolph}.\\

In this paper we investigate the problem of communicating three parameters in the absence of SRF, explicitly we assume that Alice has encoded the information in three angles $\alpha$, $\beta$ and $\gamma$ and wants to send them to Bob. We propose two different methods and compare them with regard to the efficiency of information communication per spin. In the first method (method A) Alice prepares three spins with mutual relative angles $\alpha$, $\beta$ and $\gamma$, and in the second method (method B) she uses three separate two-spin systems to send each parameter separately, see figure (\ref{picture}). We quantify the efficiency of these two methods by calculating the average information gain of Bob per spin, and we show that in method A, Bob will gain more information per spin than in method B.\\

The paper is organized as follows: in section (\ref{sec2}) we review the notion of relative and collective parameters and investigate the optimal measurement for estimating relative parameters. Then, in sections (\ref{example1}) and (\ref{example2}) we study the problem of sending three parameters in the absence of reference frame. In section (\ref{example1}) we consider the case that the parties use relative angles of spin $\frac{1}{2}-$particles and we compare the performance of methods A and B in this context. 
In section (\ref{example2}) we investigate the case that the parties use one relative angle of two qubits and two relative angles of spin-$1/2$ and spin-$j$ particles to encode three parameters and again we analyze both methods A and B in that situation. We then end the paper with a discussion.

\section{Preliminaries}\label{sec2}

Here we review the notion of relative and collective parameters as established in \cite{Rudolph}, we will then present the framework of the problem of sending information between two distant parties who do not have any shared reference frame.\\

Consider the Hilbert space of two spin-$j_{1}$ and spin-$j_{2}$ particles, namely $H_{j_{1}} \otimes H_{j_{2}}$. The relative and collective parameters of the states of $H_{j_{1}} \otimes H_{j_{2}}$ are defined with accord to their different behavior under the action of collective rotations. If one uses the labels $\theta$ and $\Omega$ to show the set of relative and collective parameters respectively, then under the action of a collective rotation, an arbitrary quantum state $\rho_{\theta,\Omega}$, transforms as:

\begin{equation}\label{collective and relative parameter}
R(\Omega^{'}) \ \rho_{\theta,\Omega} \ R(\Omega^{'})^{\dagger} = \rho_{\theta,\Omega^{'}\Omega} \ ,
\end{equation}
where $R(\Omega^{'})$ is a collective rotation with parameter $\Omega'$:\\
 \begin{equation}
 R(\Omega^{'})=R_{j_{1}}(\Omega^{'})\otimes R_{j_{2}}(\Omega^{'}),
 \end{equation}
i.e. two systems are rotated in the same manner.
Equation (\ref{collective and relative parameter}) shows that relative parameters are invariant under the action of collective rotations.
Here it must be noted that a typical parameter may be neither relative and nor collective, this is clear due to the mathematical definition of relative and collective parameters.\\

 \begin{figure}
\centering
\includegraphics[width=12cm,height=7cm,angle=0]{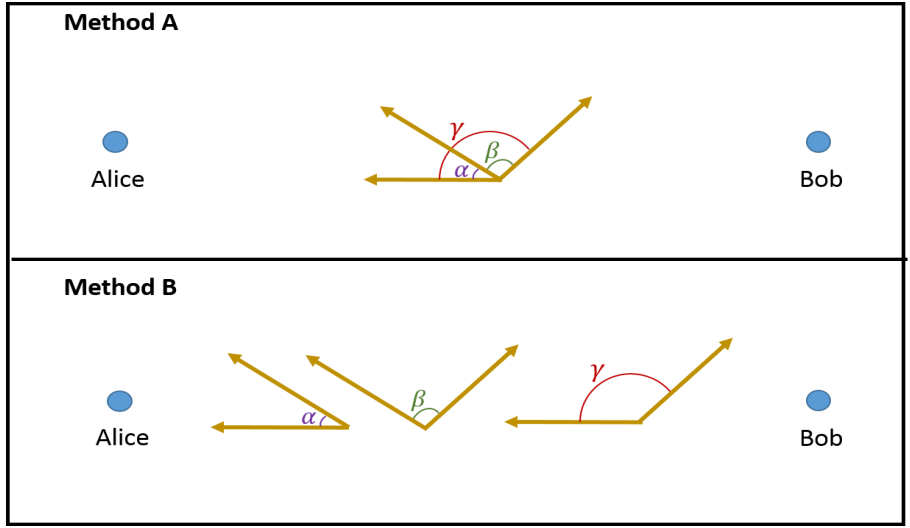}
\caption{The schematic description of two different methods of information sharing: Alice wishes to share three angles $\alpha$, $\beta$ and $\gamma$ with Bob. In method A, she sends a three-spin system to Bob, while in method B Alice sends three separate two-spin states (six spins).}
 \label{picture}
 \end{figure}
 
Now suppose that Alice and Bob do not have any shared reference frame and Alice wishes to send some information (in the form of three angles) to Bob. Due to the lack of a common frame, Alice stores the information in three relative parameters of some spins and sends them to Bob, i.e. Alice prepares the state $\rho_{\alpha,\beta,\gamma}$ (with $\alpha$, $\beta$ and $\gamma$  being relative parameters) and sends that to Bob who wishes to read the information by suitable measurements. 

The most general measurement that can be performed by Bob is a POVM represented by the elements $\left\lbrace E_\lambda\right\rbrace $. 
By obtaining the measurement result $\lambda$, Bob updates his prior knowledge of the parameters from prior probability distribution $P(\alpha,\beta,\gamma)$ to the posterior distribution $P(\alpha,\beta,\gamma|\lambda)$ by using Bayesian approach \cite{Bayesian}:
 \begin{equation}\label{posterior distribution}
 P(\alpha,\beta,\gamma\vert\lambda)=\frac{tr(E_\lambda \rho _{\alpha,\beta,\gamma})P(\alpha,\beta,\gamma)}{P(\lambda)},
 \end{equation}
 where
 \begin{equation}\label{p-lambda}
 P(\lambda)=\int tr(E_\lambda \rho _{\alpha,\beta,\gamma})P(\alpha,\beta,\gamma) d\alpha d\beta d\gamma.
 \end{equation}

The efficiency of this information sharing task, can be quantified by the average relative information between prior and posterior probability distributions, specifically
 \begin{equation}\label{I-avg}
 I_{avg} = \sum_{\lambda} p(\lambda) I_\lambda,
 \end{equation}
 where
 \begin{equation}\label{I-lambda}
 I_\lambda=\int P(\alpha,\beta,\gamma\vert\lambda) \log_2{[\frac{P(\alpha,\beta,\gamma\vert\lambda)}{P(\alpha,\beta,\gamma)}]} d\alpha d\beta d\gamma.
 \end{equation}
$I_{avg}$ can be interpreted as the average information gained by Bob. Bob then aims to perform the POVM which maximizes $I_{avg}$. For the case where Alice sends only one relative parameter to Bob (she sends two spins with relative angle $\alpha$), it has been shown in \cite{Rudolph} that the optimal measurement is the total spin measurement.
The argument of \cite{Rudolph} is also valid for a quantum state with several relative parameters, like $\rho_{\alpha,\beta,\gamma}$, and hence the optimal measurement of our problem is also the total spin measurement.\\

In the following we will present two different strategies that Alice can use to send three angles $\alpha$, $\beta$ and $\gamma$ to Bob. The first strategy is to use the relative angles of spin-$\dfrac{1}{2}$ particles and the second is to use both spin-$\dfrac{1}{2}$ and spin-$j$ particles. For any strategy, Alice can proceed by using two different methods, method A in which Alice uses three relative angles of three spin directions, or method B in which Alice uses three separate angles of three separate spin pairs to encode the information (see figure (\ref{picture})). It should be noted that the number of spins used in methods A and B are different, i.e. three spins for method A and six ones for method B. Hence, in order to have a comprehensive and more precise comparison of these methods, we define the average information gain per spin:
\begin{equation}\label{average information per spin}
\textbf{i}:=\dfrac{I_{avg}}{N},
\end{equation}
where $N$ is the number of spins used in a method. A more efficient method is the one with higher value of $\textbf{i}$.

\section{Using relative angles of spin$-\dfrac{1}{2}$ particles}\label{example1}

In this section we explore the case that Alice uses spin $-\dfrac{1}{2}$ particles for sending the angles $\alpha$, $\beta$ and $\gamma$ and we compare methods A and B in this framework. We first calculate the information gain of each method.\\

\textbf{Method A}: Alice prepares the three-qubit state 

\begin{equation}\label{qubitstate}
|\psi_{\alpha,\beta,\gamma}\rangle=|\hat{n}\rangle\otimes|\hat{m}\rangle\otimes|\hat{r}\rangle,
\end{equation}
where $|\hat{a}\rangle$ is the eigenvector of $\vec{\sigma}.\hat{a}$ with positive eigenvalue, $\vec{\sigma}=(\sigma_x , \sigma_y , \sigma_z)$ is the vector of Pauli matrices, $\alpha=\cos^{-1}(\hat{m}.\hat{n})$, $\beta=\cos^{-1}(\hat{m}.\hat{r})$ and $\gamma=\cos^{-1}(\hat{n}.\hat{r})$.
\\

She then sends the qubits to Bob who wishes to acquire the most possible information about $\alpha$, $\beta$ and $\gamma$. Bob performs his optimal measurement which is the total spin measurement of all three qubits \cite{Rudolph}. This optimal measurement is a three outcome POVM: $\left\lbrace \Pi_{\frac{1}{2}^{'}},\Pi_{\frac{1}{2}},\Pi_{\frac{3}{2}}\right\rbrace $, that is due to the fact that the joint Hilbert space of three qubits decomposes into the direct sum of three subspaces, i.e. the subspaces of total angular momentum:
\begin{equation}\label{qubitdirectsum}
 \dfrac{1}{2}\otimes \dfrac{1}{2}\otimes \dfrac{1}{2}=(0 \oplus 1)\otimes\dfrac{1}{2}=\dfrac{1}{2}^{'}\oplus\dfrac{1}{2}\oplus\dfrac{3}{2}.
\end{equation}

To clarify the notation, we note that the joint Hilbert space of three qubits has two subspaces with total spin equal to $\dfrac{1}{2}$, and we show these subspaces with  $\dfrac{1}{2}$ and $\dfrac{1}{2}^{'}$. The term $\dfrac{3}{2}$ also stands for the subspace with total spin equal to $\dfrac{3}{2}$. Having in mind this notation, the POVM element $\Pi_{k}$ is the projective operator on the subspace which is characterized with $k$.\\

It will be more convenient to calculate the probabilities of each outcome, if we expand POVM elements in terms of Pauli matrices. By a simple trick and straightforward calculations (See Appendix) we find:

\begin{eqnarray}\label{expansions}
  \Pi_{\frac{1}{2}^{'}}&=&\dfrac{1}{4}\left[ I \otimes I \otimes I - \vec{\sigma}_1.\vec{\sigma}_2 \right],\cr\cr
  \Pi_{\frac{3}{2}}&=&\dfrac{1}{2} I \otimes I \otimes I + \dfrac{1}{6}\left[ \vec{\sigma}_1. \vec{\sigma}_2 + \vec{\sigma}_2.\vec{\sigma}_3 + \vec{\sigma}_1.\vec{\sigma}_3\right] ,\cr\cr
  \Pi_{\frac{1}{2}}&=&I \otimes I \otimes I - \Pi_{\frac{3}{2}} -  \Pi_{\frac{1}{2}^{'}},
 \end{eqnarray}
where $\vec{\sigma}_i$ is the $\vec{\sigma}$-vector of particle $i$. Consequently, by considering $P(\alpha,\beta,\gamma)$ to be a uniform prior probability distribution, we calculate the conditional probabilities of each outcome as
\begin{eqnarray}
P(\Pi_{\frac{1}{2}^{'}}|\alpha,\beta,\gamma)&=&tr(\Pi_{\frac{1}{2}^{'}}\rho_{\alpha,\beta,\gamma})=\dfrac{1}{4}-\dfrac{1}{4}\cos(\alpha),\cr\cr
P(\Pi_{\frac{3}{2}}|\alpha,\beta,\gamma)&=&tr(\Pi_{\frac{3}{2}}\rho_{\alpha,\beta,\gamma})=\dfrac{1}{2}+\dfrac{1}{6}\left(\cos(\alpha)+\cos(\beta)+\cos(\gamma)\right),\cr\cr
P(\Pi_{\frac{1}{2}}|\alpha,\beta,\gamma)&=&tr(\Pi_{\frac{1}{2}}\rho_{\alpha,\beta,\gamma})=\dfrac{1}{4}+\dfrac{1}{12}\cos(\alpha)-\dfrac{1}{6}\cos(\beta)-\dfrac{1}{6}\cos(\gamma),
\end{eqnarray}
which implies
\begin{equation}
P(\Pi_{\frac{1}{2}^{'}})=P(\Pi_{\frac{1}{2}})=\dfrac{1}{4}, \h
P(\Pi_{\frac{3}{2}})=\dfrac{1}{2}.
\end{equation}
Now, the posterior probability distribution can be obtained with accord to equation (\ref{posterior distribution}) and finally the average information gain of each outcome will be calculated numerically by using equation (\ref{I-lambda})
\begin{equation}
I_{\Pi \frac{1}{2}^{'}}=0.27, \h
I_{\Pi \frac{3}{2}}=0.07, \h
I_{\Pi \frac{1}{2}}=0.24,
\end{equation}
which yields
\begin{equation}
 I_{avg}=0.17. 
\end{equation}
So the average information gain per spin for method A is  $\textbf{i}=0.056$.\\

\textbf{Method B}: As illustrated in figure (\ref{picture}), another way for transferring three angles $\alpha$, $\beta$ and $\gamma$ is to send three separate pairs of qubits instead of three qubits. Alice prepares three separate pairs of qubits, encodes the information in the relative angle of each pair and sends them to Bob. The relative angle of the first, second and third pairs are denoted by $\alpha$, $\beta$ and $\gamma$ respectively. The six-qubit state can then be written as:
\begin{equation}
\rho=\rho_1\otimes\rho_2\otimes\rho_3,
\end{equation} 
where $\rho_i$ describes the state of $i-$th pair.
Bob performs his optimal measurement on each pair separately. The optimal measurement for retrieving the relative angle of two qubits is the total spin measurement \cite{Rudolph}  described by two projectors $\Pi_0=|\psi^-\rangle\langle\psi^-|$ and $\Pi_1=I-|\psi^-\rangle\langle\psi^-|$, where $|\psi^-\rangle$ is the singlet state. Therein, the POVM elements on the six qubits can be written as:
\begin{equation}
\Pi_{ijk}=\Pi_i \otimes \Pi_j \otimes \Pi_k, \h
i,j,k = {0,1},
\end{equation}
and the conditional probabilities of each outcome, given $\alpha$, $\beta$ and $\gamma$ are equal to
\begin{equation}
P(\Pi_{ijk}|\alpha,\beta,\gamma)=tr(\Pi_{ijk}\rho)=P(\Pi_i|\alpha) P(\Pi_j|\beta) P(\Pi_k|\gamma).
\end{equation}
Hence, by straightforward calculations based on equations (\ref{posterior distribution}) and (\ref{p-lambda}), one sees that the posterior probability distribution of the three angles has a product form:
\begin{equation}\label{ppp}
P(\alpha,\beta,\gamma| \Pi_{ijk})=P(\alpha|\Pi_i) P(\beta|\Pi_j)P(\gamma|\Pi_k).
\end{equation}
By inserting (\ref{ppp}) into (\ref{I-avg}), we see that the average information gain of method B has an additive behavior:
\begin{equation}
I_{avg}=\bar{I}_{\alpha}+\bar{I}_{\beta}+\bar{I}_{\gamma},
\end{equation}
where $\bar{I}_{\alpha}$ is the average information gain of $\alpha$ when Bob performs a total spin measurement on the first qubit pair, $\bar{I}_{\beta}$ and $\bar{I}_{\gamma}$ have also similar definitions. It is obvious that $\bar{I}_{\alpha}=\bar{I}_{\beta}=\bar{I}_{\gamma}$.
For initial uniform probability distribution, it has been shown in \cite{Rudolph} that the maximum accessible average information gain of the relative angle between two qubits is $0.08$, it implies that $I_{avg}=0.24$ and therefore the average information gain per spin is $\textbf{i}=0.04$ .\\

We see that when Alice uses qubits to encode the information, the average information gain per spin in method A, $0.05$ , is larger than that of method B, $0.04$. So in this case method A is more efficient than method B.\\

The superiority of method A to method B for qubits, creates a very important class of open problems. One can ask does method A lead to higher value of information gain per spin than method B, even for sending more than three parameters, and even by using spin$-j$ particles instead of spin$-\dfrac{1}{2}$ particles? The importance of the issue is due to the lower number of particles that are used in method A\footnote{We will discuss more explicitly in the discussion section.}, and it motivates the investigation of problem from different perspectives. In the next section, we replace some of spin$-\dfrac{1}{2}$ particles with spin$-j$ ones and we compare the efficiency of the methods.

\section{Using spin$-j$ and spin$-\dfrac{1}{2}$ particles}\label{example2}
In this section we consider the other strategy of Alice and compare methods A and B in that context. Again we suppose that Alice wants to transfer some information to Bob in the absence of a common reference frame, and due to lack of a common frame, she sends the information through the relative angles of the spins. As the second strategy for sending three angles, Alice encodes $\alpha$ in the relative angle of two qubits, but she encodes $\beta$ and $\gamma$ in the relative angles of spin$-\dfrac{1}{2}$ and spin$-j$ particels. In the following, we will calculate the efficiency of methods A and B, for this strategy.\\

\textbf{Method A:} Here, Alice sends two spin$-\dfrac{1}{2}$ and one spin$-j$ particles to Bob. The quantum state is represented as:
\begin{equation}\label{initial state}
|\chi'\rangle= |\hat{n}\rangle\otimes|\hat{m}\rangle\otimes|\hat{r}_j\rangle,
\end{equation}
where $|\hat{n}\rangle$ and $|\hat{m}\rangle$ are defined as in equation (\ref{qubitstate}) and $|\hat{r}_j\rangle$ is a SU(2) coherent state (the eigenstate of $\vec{J}.\hat{r}$ associated with the maximum eigenvalue). Bob seeks to gain some information about the relative angles $\alpha=\cos^{-1}(\hat{n}.\hat{m})$, $\beta=\cos^{-1}(\hat{n}.\hat{r})$ and $\gamma=\cos^{-1}(\hat{m}.\hat{r})$. Again he performs the optimal total spin measurement on the three spins.  The joint Hilbert space of two spin$-\dfrac{1}{2}$ and one spin$-j$ can be decomposed to the direct sum of four irreducible subspaces:
 \begin{equation}
 \dfrac{1}{2}\otimes\dfrac{1}{2}\otimes j =(j-1) \oplus j \oplus(j+1) \oplus j',
 \end{equation} 
where we have used the same notation as equation (\ref{qubitdirectsum}). Therein the three  particle POVM, performed by Bob, has four elements: $\left\lbrace \Pi_{j-1},\Pi_j,\Pi_{j+1},\Pi_{j'}\right\rbrace $.
To calculate the probabilities of each outcome, one may follow the same steps as section (\ref{example1})  which were presented in the Appendix, but due to the presence of spin $j$, that calculation will be very complicated. Hence we take another route and expand the initial state in the total spin basis. With accord to uniformity of prior probability distributions, and also the integration in equation (\ref{I-lambda}), without loss of generality, we can consider $\hat{r}$ to be along the $\textbf{z}$ axis and $\hat{m}$ to lie in $\textbf{z-x}$ plane of Alice, i.e. we use the parametrization $\hat{n}=(\sin{\frac{\theta}{2}} \cos{\phi}, \sin{\frac{\theta}{2}} \sin{\phi}, \cos{\frac{\theta}{2}})$ and $\hat{m}=(\sin{\frac{\theta'}{2}} , 0 , \cos{\frac{\theta'}{2}})$.
Applying this notation, the initial state (\ref{initial state}) can be rewritten as:
\begin{equation}\label{initial state 2}
|\chi'\rangle=\left(\cos(\dfrac{\theta}{2})|0\rangle+\sin(\dfrac{\theta}{2}) e^{i\phi}|1\rangle\right)\otimes\left(\cos(\dfrac{\theta'}{2})|0\rangle+\sin(\dfrac{\theta'}{2})|1\rangle\right) \otimes |\hat{z}_j\rangle,
\end{equation} 
where $|0\rangle$ and $|1\rangle$ are the eigenvectors of $\vec{\sigma}.\hat{z}$. Now by using the Clebsh-Gordon coefficients, we can expand all terms of (\ref{initial state 2}) in the basis of total spin eigenvectors: 

\begin{eqnarray}
  |0,0,\hat{z}_j\rangle &=& |j+1,j+1\rangle, \cr\cr
 |0,1,\hat{z}_j\rangle &=& \dfrac{1}{\sqrt{2j+1}}|j+1,j\rangle - \sqrt{\dfrac{2j+1}{2j+2}}|j,j\rangle, \cr\cr
 |1,0,j\hat{z}\rangle &=&\dfrac{1}{\sqrt{2j+2}}|j+1,j\rangle+\dfrac{1}{\sqrt{(2j+1)(2j+2)}}|j,j\rangle-\sqrt{\dfrac{2j}{2j+1}}|j',j'\rangle, \cr\cr
 |1,1,j\hat{z}\rangle &=& \dfrac{1}{\sqrt{(2j+1)(j+1)}}|j+1,j-1\rangle-\sqrt{\dfrac{j}{(2j+1)(j+1)}}|j,j-1\rangle\cr
 & & -\dfrac{1}{\sqrt{2j+1}}|j',j'-1\rangle+\sqrt{\dfrac{2j-1}{2j+1}}|j-1,j-1\rangle.
\end{eqnarray}
Inserting these equations in (\ref{initial state 2}), the conditional probabilities of each outcome, i.e.  $P(\Pi_i|\theta, \phi, \theta')= tr(\Pi_i|\chi'\rangle\langle\chi'|)$, can be calculated easily and then with accord to equation (\ref{p-lambda}) we find:
\begin{equation}
\begin{split}
P(\Pi_{j'})=\dfrac{1}{4}, \h
P(\Pi_{j+1})=\dfrac{2j+3}{8j+4}, \\
P(\Pi_j)=\dfrac{1}{4}, \h
P(\Pi_{j-1})=\dfrac{2j-1}{8j+4}.
\end{split} 
\end{equation}

\begin{figure}
\centering
\includegraphics[width=12cm,height=8cm,angle=0]{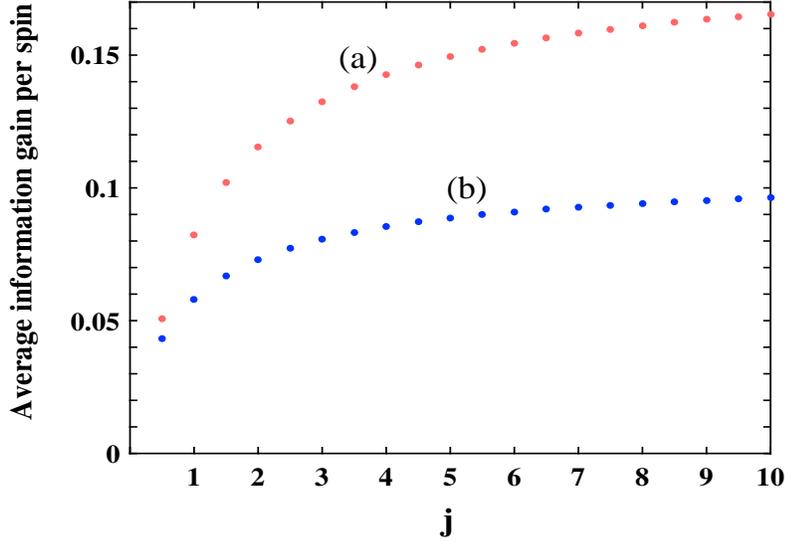}
\caption{Average information gain per spin of method A (curve (a)) and method B (curve (b)) for different values of $j$.}
\label{comparison}
\end{figure}

The posterior probabilities will be obtained by inserting the above equations and conditional probabilities in equation (\ref{posterior distribution}). Henceforth we numerically  calculate the average information gain per spin as a function of $j$, and the results are presented in curve (a) of figure (\ref{comparison}). Figure (\ref{comparison}) shows that the average information gain per spin is a monotonically increasing function of $j$ and it reaches $\textbf{i}=\frac{0.55}{3}=0.18$ for large values of $j$. It should be noted that in this limit, all measurement outcomes have the same information gain (i.e. $I_{\Pi j'}=I_{\Pi j+1}=I_{\Pi j}=I_{\Pi j-1}=0.55$) which can be understood from the equality of dimensions of total spin subspaces as  $j\rightarrow\infty$.\\

\textbf{Method B:} Here we explore the efficiency of method B for our second strategy. That is when Alice sends three parameters to Bob separately instead of sending them simultaneously.
To have a comprehensive comparison with method A, we suppose that Alice uses a pair of spin$-\frac{1}{2}$ particles to send the angle $\alpha$ and for sending $\beta$ and $\gamma$, she uses two spin pairs, each of which is composed of one spin$-\frac{1}{2}$ and one spin$-j$ particle. 
On the other side, Bob performs the optimal measurement on his individual pairs to get the most accessible information about the relative angles. As it was shown in section (\ref{example1}), the average information gain of method B is additive:
\begin{equation}
I_{avg}=\bar{I}_{\alpha}+\bar{I}_{\beta}+\bar{I}_{\gamma}.
\end{equation}
Here $\bar{I}_{\beta}=\bar{I}_{\gamma}\neq\bar{I}_{\alpha}$ and all these quantities can be obtained numerically as functions of $j$, by following the same steps as in \cite{Rudolph}. Curve (b) of figure (\ref{comparison}) shows $\textbf{i}=\frac{\bar{I}_{avg}}{6}$ for method B, it is clear that the average information gain per spin, \textbf{i}, is again a monotonically increasing function which tends to the value $0.10$ as  $j\rightarrow\infty$.\\

As seen from figure (\ref{comparison}), for $j=\frac{1}{2}$ the results of previous section are retrieved, and method A has a better performance than B for all values of $j$. This is valuable since method A uses less particles than method B (here 3 ones in comparison to 6 ones).

\section{Discussion}

In the absence of a Shared Reference Frame, encoding the information in relative parameters of a quantum system can be a beneficial way of communication. 
Here we investigate two different methods (A and B) for sharing three parameters and compare their efficiency. 
More generally, to share $N$ parameters, the parties can have two different methods. In method A, $N$ parameters can be encoded in mutual relative angles of $M$ spin directions ( $\begin{pmatrix} M \\ 2 \end{pmatrix} =N$), while in method B, the angles of $N$ separate spin pairs can be used for information sharing. Evidently method A uses exponentially less particles than B, for large values of $N$. 
We investigated this problem for $N=3$ and within two different strategies. 
One strategy employs spin$-\frac{1}{2}$ particles and the other uses both spin$-\frac{1}{2}$ and spin $j$ particles. Our results show that, method A not only uses less particles, but also is more efficient than method B, i.e. the average information gain per spin is higher for method A. We conjecture that method A is more efficient than method B even for sending more than three spins, and even by using spin$-j$ particles instead of spin$-\dfrac{1}{2}$ ones, but the problem is still open and it is of great interest due to the exponential decrease of the required spins. 


\section*{Appendix}
Here we present the mathematics that we used to derive equation (\ref{expansions}). We find the expansions (\ref{expansions}) by solving a system of three linear equations.\\

The subspace $J=\dfrac{1}{2}^{'}$ is an irreducible subspace which corresponds to $0\otimes\dfrac{1}{2}$. With accord to the fact that $|\psi^{-}\rangle=\dfrac{1}{\sqrt{2}}(|01\ra-|10\ra)$ is the only two-qubit state with $J_{total}=0$, the equation which determines $\Pi_{\frac{1}{2}^{'}}$ is written as
\begin{equation}\label{equation1}
\Pi_{\frac{1}{2}^{'}}=|\psi^{-}\rangle\langle\psi^{-}| \otimes I=\dfrac{1}{4}\left[ I \otimes I \otimes I - \vec{\sigma}_1 . \vec{\sigma}_2 \right], 
\end{equation}
where the index $i$, in $\vec{\sigma}_i$, shows the particle number and $\vec{\sigma}_i . \vec{\sigma}_j=\sigma_{ix} \otimes \sigma_{jx} + \sigma_{iy} \otimes \sigma_{jy} + \sigma_{iz} \otimes \sigma_{jz} $. 
Another equation is derived from the fact that all projectors sum up to identity operator:
\begin{equation} \label{equation2}
\Pi_{\frac{1}{2}^{'}}+\Pi_{\frac{1}{2}}+\Pi_{\frac{3}{2}}=I \otimes I \otimes I.
\end{equation}
Finally the last equation can be obtained by expanding the total spin operator $\vec{J}.\vec{J}$ in terms of the projectors $\Pi_{\frac{1}{2}'}$, $\Pi_{\frac{1}{2}}$ and $\Pi_{\frac{3}{2}}$. Subspaces $J=\dfrac{1}{2}'$, $J=\dfrac{1}{2}$ and $J=\dfrac{3}{2}$ are eigenspaces of  $\vec{J}.\vec{J}$ with the eigenvalue $J(J+1)$, hence

\begin{equation}\label{equation3}
\vec{J}.\vec{J}= \dfrac{3}{4}\Pi_{\frac{1}{2}^{'}}+\dfrac{3}{4}\Pi_{\frac{1}{2}^{'}}+\dfrac{15}{4}\Pi_{\frac{3}{2}}.
\end{equation}\\
One can now solve linear equations (\ref{equation1}), (\ref{equation2}) and (\ref{equation2}) for  $\Pi_{\frac{1}{2}'}$, $\Pi_{\frac{1}{2}}$ and $\Pi_{\frac{3}{2}}$ to derive equation (\ref{expansions}).

{}

\end{document}